\documentclass[sigconf]{acmart}
\usepackage{graphicx}
\usepackage{subcaption} 
\usepackage{enumitem}
\usepackage{multirow}

\AtBeginDocument{%
  }

\setcopyright{acmlicensed}
\copyrightyear{2026}
\acmYear{2026}
\setcopyright{cc}
\setcctype{by-nc-nd}
\acmConference[CHI '26]{Proceedings of the 2026 CHI Conference on Human Factors in Computing Systems}{April 13--17, 2026}{Barcelona, Spain}
\acmBooktitle{Proceedings of the 2026 CHI Conference on Human Factors in Computing Systems (CHI '26), April 13--17, 2026, Barcelona, Spain}
\acmPrice{}
\acmDOI{10.1145/3772318.3791978}
\acmISBN{979-8-4007-2278-3/2026/04}

\begin{document}

%%
%% The "title" command has an optional parameter,
%% allowing the author to define a "short title" to be used in page headers.
\title{Living with Data: Exploring Physicalization Approaches to Sedentary Behavior Intervention for Older Adults in Everyday Life}

%%
%% The "author" command and its associated commands are used to define
%% the authors and their affiliations.
%% Of note is the shared affiliation of the first two authors, and the
%% "authornote" and "authornotemark" commands
%% used to denote shared contribution to the research.
\author{Siying Hu}
% \affiliation{%
%   \institution{City University of Hong Kong}
%   \country{Hong Kong SAR}
%   }
\affiliation{%
  \institution{The University of Queensland}
  \city{Brisbane}
  \state{Queensland}
  \country{Australia}
}
\email{siying.hu@uqconnect.edu.au}
\orcid{0000-0002-3824-2801}

\author{Zhenhao Zhang}
\affiliation{%
  \institution{City University of Hong Kong}
  \country{Hong Kong SAR}
  }
\email{zhenhao.research@gmail.com}
\orcid{0009-0001-9242-9618}

%%
%% By default, the full list of authors will be used in the page
%% headers. Often, this list is too long, and will overlap
%% other information printed in the page headers. This command allows
%% the author to define a more concise list
%% of authors' names for this purpose.
\renewcommand{\shortauthors}{Hu and Zhang}

%%
%% The abstract is a short summary of the work to be presented in the
%% article.
\begin{abstract}
Sedentary behavior is a critical health risk for older adults. Although digital interventions are widely available, they primarily rely on screen-based notifications that can feel clinical or cognitively demanding, and are thus often ignored over time. This paper presents a three-phase Research through Design methodology to explore data physicalization approaches that ambiently represent sedentary data patterns using decor artifacts in older adults’ homes. These artifacts transformed abstract data into aesthetic, evolving forms that became part of the domestic landscape. Our research revealed how these physicalizations fostered self-reflection, family conversations, and encouraged active lifestyles. We demonstrate how qualities like aesthetic ambiguity and slow revelation can empower older adults, fostering a reflective relationship with their well-being. Ultimately, we argue that creating data physicalizations for older adults necessitates a shift from merely informing users to enabling them to live with and through their data.
\end{abstract}

%%
%% The code below is generated by the tool at http://dl.acm.org/ccs.cfm.
%% Please copy and paste the code instead of the example below.
%%


%%
%% Keywords. The author(s) should pick words that accurately describe
%% the work being presented. Separate the keywords with commas.

%% A "teaser" image appears between the author and affiliation
%% information and the body of the document, and typically spans the
%% page.
\begin{teaserfigure}
  \includegraphics[width=\textwidth]{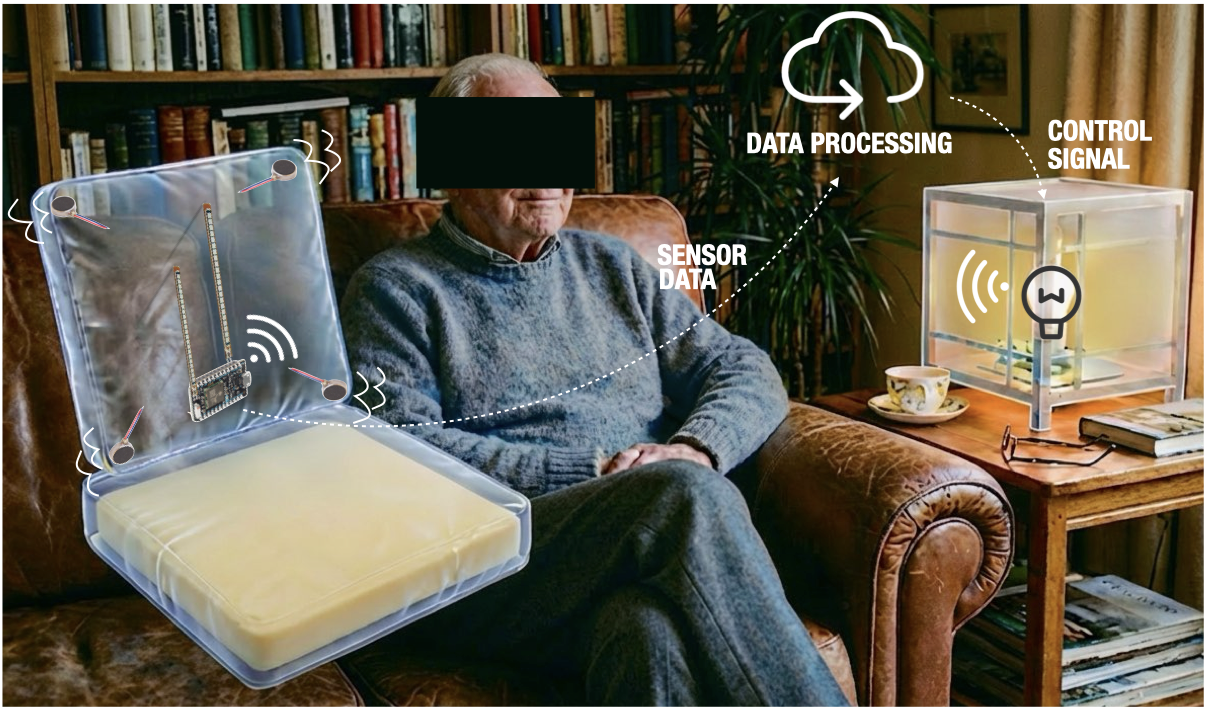}
  \caption{The \textit{CushLumi} system embedded in an older adult’s living room: flexible sensors inside a familiar seat cushion capture sedentary behavior and send sensor data to a cloud-based processor, which in turn drives an ambient lamp that gradually changes its light to physicalize sitting patterns. Rather than issuing screen-based alerts, the system weaves personal sedentary data into the domestic atmosphere, gently inviting reflection and movement within everyday life.}
  \Description{the \textit{CushLumi} embedded in an older adult’s living room: flexible sensors inside a familiar seat cushion capture sedentary behavior and send sensor data to a cloud-based processor, which in turn drives an ambient lamp that gradually changes its light to physicalize sitting patterns. Rather than issuing screen-based alerts, the system weaves personal sedentary data into the domestic atmosphere, gently inviting reflection and movement within everyday life.}
  \label{fig:teaser}   
\end{teaserfigure}

%%
%% This command processes the author and affiliation and title
%% information and builds the first part of the formatted document.
\maketitle
\begin{figure*}[t]
    \centering
    \includegraphics[width=1\linewidth]{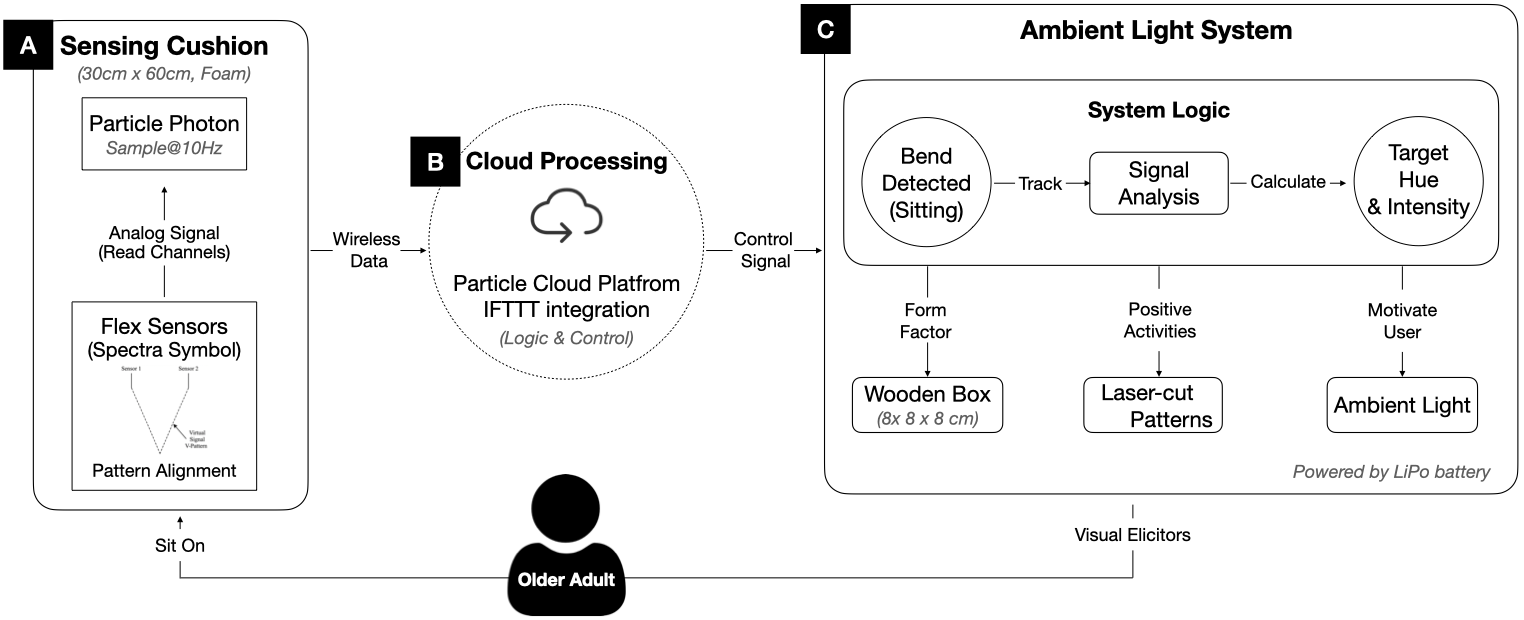}
    \caption{The schematic diagram of the \textit{CushLumi} System: A Cloud-Based Smart Home Lighting System Driven by Cushion Sensors. The system consisted of three parts: A) a Sensing Cushion and C) an Ambient Light lamp, which are the two physical artefacts users directly engage with, and C) a cloud‑based processing component that acts as an invisible bridge, receiving sit/stand events from the cushion and driving the lamp’s gradual light changes and celebratory animations. Whenever a user sat on the cushion, its deformation was detected, and the system began tracking the duration of one's sitting. This information was then used to change the lamp's hue and light intensity to motivate the user to stand up and move around.}
    \label{fig:schematicdiagram}
\end{figure*}
\section{Introduction}

A suite of health and wellness technologies has been developed to help people monitor their daily activity, sleep, and sedentary behavior \cite{gao2019emerging,sugiyama2008joint,owen2010sedentary}. These technologies include wearable trackers and smartwatches that visualize one's steps, heart rate, and sitting time via dashboards and notifications \cite{Reinschluessel2025,tremblay2010physiological}. They also include home-based systems that provide ambient cues and persuasive prompts, such as light- or sound-based reminders to encourage movement breaks \cite{10.1145/3640794.3665561,7927391,Yu2025}. Such systems have been designed for contexts that value numerical precision and individual self-optimization, emphasizing screen-based feedback, charts, and goals as primary interaction paradigms \cite{info:doi/10.2196/54375, info:doi/10.2196/25171}.

However, with the aging population and the widespread diffusion of consumer health devices, these systems are located in households where needs and norms often differ from their original design contexts. Older adults, for example, often prioritize simplicity, comfort, and emotional ease over granular analytics \cite{app15105458}. The home is also a social and aesthetic setting where devices are expected to blend in with furniture, routines, and rhythms rather than demand active attention \cite{Ghorayeb2023,Yu2025}. This prompts reflection on the suitability of screen- and number-centric tools for people who may have increased cognitive load, fine motor challenges, or anxiety from constant alerts \cite{Sagga2025, MILES2024264}. For sedentary behavior in particular, the gap between raw sensor readings and immediately meaningful cues in everyday life remains underexplored \cite{info:doi/10.2196/54375}.

To address this gap, this research explores the physicalizations of ambient data for sedentary behavior in older adults’ homes. We report on a multi-stage Research through Design (RtD) process that employed in-situ observations, semi-structured interviews, and a prototype deployment in end users' homes. Our research sought to understand:
\begin{itemize}
    \item \textbf{RQ1:} How can abstract data on sedentary behavior be translated into a physical form that is accessible to older adults?
    \item \textbf{RQ2:} How can an ambient data physicalization approach differ from traditional screen-based digital reminders?
\end{itemize}

As a result of our research, we make three contributions. 
First, we extend prior work on data physicalization by understanding the potential and perceptions of data physicalization by evaluating its application specifically for older adults within domestic environments. These findings were identified via a three-phase Research through Design process that culminated in an in-situ deployment of an ambient data physicalization system (i.e., the \textit{CushLumi}) in older adults’ homes. 
Second, we operationalize the notion of \textit{living with data} by proposing a perspective shift that allowed for personal sedentary data to be integrated into a physical model that shaped environmental ambiance and felt atmospheres (rather than dashboards or notification alarms). 
Third, we distill our findings into design implications for future tangible health technologies, highlighting how the decoupling of sensing from display, the use of positive metaphors, and the integration of feedback into familiar domestic objects can preserve the richness of personal histories while offering non-disruptive, experiential feedback that supports older adults’ autonomy and ongoing reflection.

\section{Related Work}
A substantial body of research has demonstrated the importance of supporting gentle, situated forms of engagement with everyday information via ambient technologies. Such research has integrated cues into the periphery of daily life, allowing users to remain aware without being interrupted or cognitively overloaded \mbox{\cite{10.1145/3670653.3670681,10.1093/iwc/iwae054}}. In parallel, research on data physicalizations has shown how transforming abstract information into tangible and aesthetic forms can slow interaction, foster curiosity through ambiguity and gradual revelation, and create opportunities for reflective engagement beyond screen-based interfaces \mbox{\cite{Bae2022,9200790,miao2025physlens,10.1145/3706598.3713267}}. Despite these complementary strengths, relatively little research has explored how ambient qualities can be embedded within data physicalization practices to represent everyday bodily states, such as patterns of sitting or inactivity, or how such hybrid representations might become enduring, meaningful elements within older adults’ domestic environments.

Supporting older adults' well-being ultimately requires interventions that can foster sustained, long-term changes in everyday activity patterns \mbox{\cite{Madigan2024, Ominyi2024}}. However, existing behavior-change technologies have primarily relied on wearables and mobile apps that motivate activity through monitoring, goal setting, and notifications \mbox{\cite{info:doi/10.2196/54375}}. While such systems can be effective in the short term, engagement frequently declines over time, limiting their impact on long-term reductions in sedentary behavior \mbox{\cite{Brickwood2019}}. More recent studies have begun to explore ambient and embodied prompts that use light, sound, or physical cues to encourage brief standing or walking breaks, demonstrating that situated cues can facilitate micro-behavior changes without disrupting daily routines \mbox{\cite{10.1145/3640794.3665561,7927391,Zenscape2024}}. Many of these interventions tend to be short-term focused and implemented in controlled laboratory settings, making it unclear how well they align with the lived rhythms of everyday environments. This gap underscores the opportunity to create ambient, physicalized representations that sustain interest, invite ongoing interpretation, and gently support adaptive relationships with wellbeing data within the material fabric of everyday domestic life.

% how to considering older adults' need based on the basis 
To support subtle, long-term behavior change with ambient data physicalizations, it is crucial to understand how older adults prefer everyday behavioral information to be conveyed. Prior research found that older adults valued simplicity, comfort, and low-stress interactions rather than detailed dashboards or frequent alerts \mbox{\cite{app15105458}}. Gerontechnology studies further highlighted the importance of minimizing cognitive load and designing technologies that align with the aesthetics and rhythms of domestic life \mbox{\cite{Laukka2024,Ghorayeb2023}}. Although existing systems attempt to provide lightweight or non-intrusive visualizations for aging populations \mbox{\cite{10.1145/3613904.3642776, 10.1145/3706599.3720001}}, they often remain screen-based or reminder-driven, which is counter to the values of older adults. Considering that furniture and lighting are both influential for older adults’ physical and emotional well-being and constitute familiar \mbox{\cite{Yu2025}}, pervasive elements of domestic interiors, they offer particularly promising channels for embedding subtle, situated health-related feedback.

\section{Methodology}
To understand how sedentary behavior data might be captured, represented, and situated in older adults' everyday environments, we employed a Research through Design (RtD) methodology~\cite{10.1145/1240624.1240704,gaver2012should}. Our research had three phases: ethnographic observations (Phase 1), a poster-based discussion with diverse stakeholders (Phase 2), and a prototype deployment (Phase 3). This methodological approach was chosen because our research focuses on a problem that does not have a single optimal solution and requires the generation of new design knowledge through iterative cycles of making and reflecting. 

\begin{table*}[!htbp]
  \centering
  \small
  \setlength{\tabcolsep}{5pt}
  \renewcommand{\arraystretch}{1.1}
  \caption{Demographic overview of participants and their involvement in the three RtD phases (i.e., ethnographic observations and semi-structured interviews (Phase 1), poster-based discussion with diverse stakeholders (Phase 2), and prototype deployment and semi-structured interviews (Phase 3)). A checkmark (\checkmark) indicates that the participant took part in that phase, while a circle ($\bigcirc$) indicates that they completed a follow‑up interview in that phase.}
  \label{tab:all_participants}
  \begin{tabular}{lcccccccc} 
    \toprule
    \textbf{ID} & \textbf{Age} & \textbf{Gender} & \textbf{Role} & \textbf{Phase 1} & \textbf{Interview in Phase 1} & \textbf{Phase 2} & \textbf{Phase 3} \\
    \midrule
    P1   & 71 & Female & Older adult & \checkmark & $\bigcirc$    & --          & --         \\
    P2   & 70 & Male   & Older adult & \checkmark & $\bigcirc$    & --          & --         \\
    P3   & 71 & Male   & Older adult & \checkmark & $\bigcirc$    & \checkmark  & --         \\
    P4   & 63 & Male   & Older adult & \checkmark & $\bigcirc$    & --          & \checkmark \\
    P5   & 79 & Female & Older adult & \checkmark & --            & --      & --         \\
    P6   & 80 & Female & Older adult & \checkmark & --          & --          & --         \\
    P7   & 68 & Female & Older adult & \checkmark & --          & --          & --         \\
    P8   & 63 & Female & Older adult & \checkmark & $\bigcirc$          & --          & \checkmark \\
    P9   & 61 & Male   & Older adult & \checkmark & $\bigcirc$          & --          & \checkmark \\
    P10  & 64 & Female & Older adult & \checkmark & --          & --          & --         \\
    P11  & 64 & Male   & Older adult & \checkmark & $\bigcirc$          & --          & \checkmark \\
    P12  & 63 & Female & Older adult & \checkmark & $\bigcirc$         & --          & --         \\
    P13  & 67 & Female & Older adult & \checkmark & $\bigcirc$          & \checkmark  & --         \\
    P14  & 69 & Male   & Older adult & \checkmark & $\bigcirc$         & --          & \checkmark \\
    P15  & 71 & Female & Older adult & \checkmark & $\bigcirc$          & \checkmark  & --         \\
    \midrule
    % ---------- Experts and designers (Phase 2 only) ----------
    R1   & 45 & Female & Geriatric medicine researcher       & -- & --    & \checkmark & -- \\
    R2   & 43 & Female & Public health researcher            & -- & --          & \checkmark & -- \\
    R3   & 52 & Male   & Biomedical sciences researcher      & -- & --          & \checkmark & -- \\
    D1   & 29 & Female & Interaction designer (healthcare)   & -- & --   & \checkmark & -- \\
    D2   & 22 & Male   & Interaction designer (accessibility)  & -- & --          & \checkmark & -- \\
    D3   & 24 & Female & Interaction designer (older adults) & -- & --   & \checkmark & -- \\
    \bottomrule
  \end{tabular}
\end{table*}

\subsection{Participants}
Our RtD process involved several participants, some of whom participated in multiple phases (\autoref{tab:all_participants}). All participants were recruited from Australia, primarily through a local nursing home or nearby community centres that regularly hosted activities for older adults. Recruitment materials and staff referrals targeted individuals who self-identified as having predominantly sedentary daily routines and who were able to sit down and stand up independently without acute mobility impairments. We did not apply strict clinical criteria, instead focusing on older adults who self-reported spending substantial time sitting at home and were interested in reflecting on their daily activity patterns.

For Phase 1, we collaborated with a local nursing home to recruit 15 participants for our ethnographic observations. We then conducted semi-structured interviews with 12 of these participants after the ethnographic observations. For Phase 2, we recruited 9 stakeholders. Three were researchers specializing in geriatric medicine, public health, and biomedical sciences, another three were interaction designers with experience in healthcare and systems for older adults, and the remaining three were older adults who represented our target end users. These stakeholders had diverse perspectives and levels of prior engagement with personal data, clinical, design, and lived-experiences, similar to prior work \cite{9200790}. For Phase 3, we recruited 5 older adults with sedentary habits. Each participant hosted the \textit{CushLumi} systemin their home for a full day and then participated in semi-structured interviews to capture their lived experience using the system. Herein, we refer to older adult participants as P1–P15, expert researchers as R1–R3, and interaction designers as D1–D3 (\autoref{tab:all_participants}).

\subsection{Data Physicalization Design Concepts and the \textit{CushLumi} Prototype}

Insights from Phase 1 led us away from a smartphone‑based intervention toward feedback that could be quietly embedded in everyday surroundings. Participants reported disliking complex apps and ``nagging'' notifications, and cushions repeatedly appeared as mundane but central artefacts in long sitting episodes. At the same time, prior work highlights cushions and lighting as familiar, pervasive elements of older adults’ domestic interiors and as promising channels for low‑priority ambient feedback \cite{wisneski1998ambient,7927391,mclaughlin2020designing,feine2025integrating}. Building on this, we developed \textit{CushLumi} as a two‑part, hybrid data physicalization system: a sensing cushion that captures sedentary behaviour, and a separate ambient lamp that materializes this data in the home.

\begin{figure*}[htbp]
    \centering
    \includegraphics[width=1\linewidth]{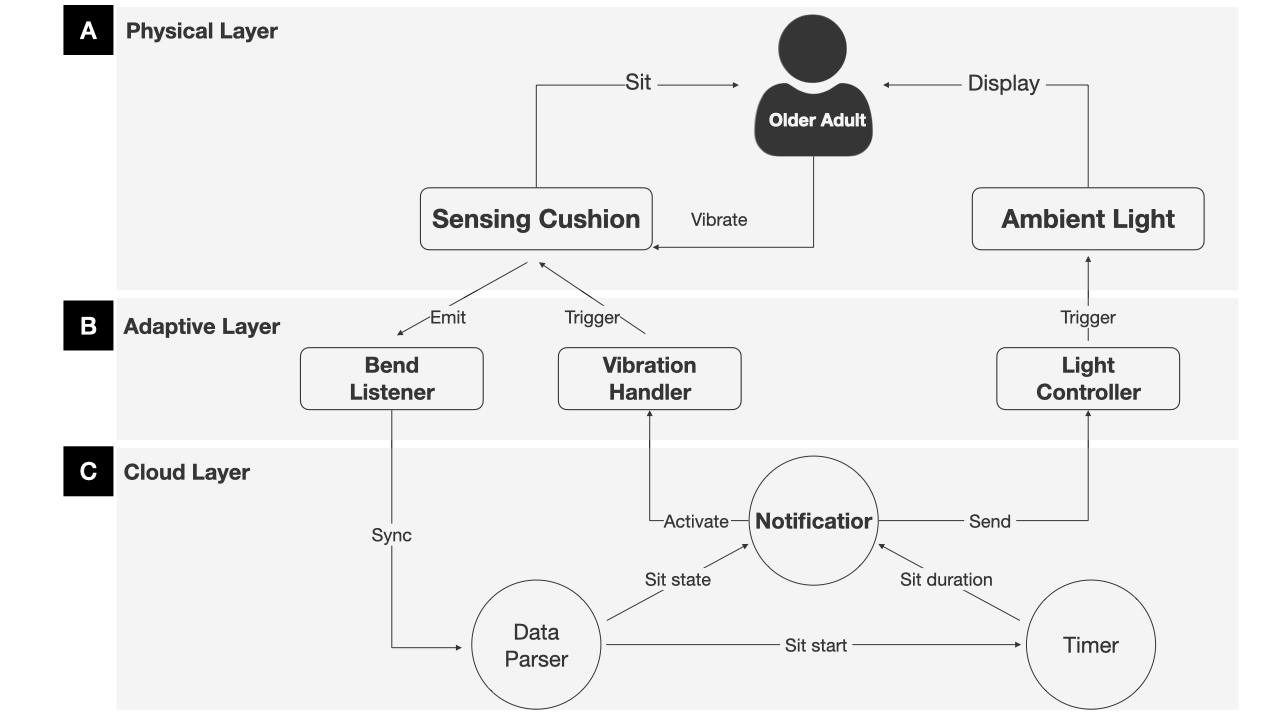}
    \caption{The the technical implementation of the \textit{CushLumi} System. At A) the physical layer, a sensing cushion detects when an older adult sits or stands and streams bend events to the cloud, while a separate ambient lamp displays the current sedentary state. B) The adaptive layer translates raw bend events into higher‑level sitting states: a bend listener detects cushion deformation, a timer tracks sit duration, and a notifier evaluates when gentle prompts or celebrative feedback should be issued. C) In the cloud layer, parsed sensor data are synchronized and used to drive a light controller that gradually increases the lamp’s brightness and shifts its colour from warm white toward red as uninterrupted sitting accumulates, and then briefly plays a colourful animation when the person stands.}
    \label{fig:technicalimplementation}
\end{figure*}

\begin{figure*}[htbp]
    \centering
    \includegraphics[width=1\linewidth]{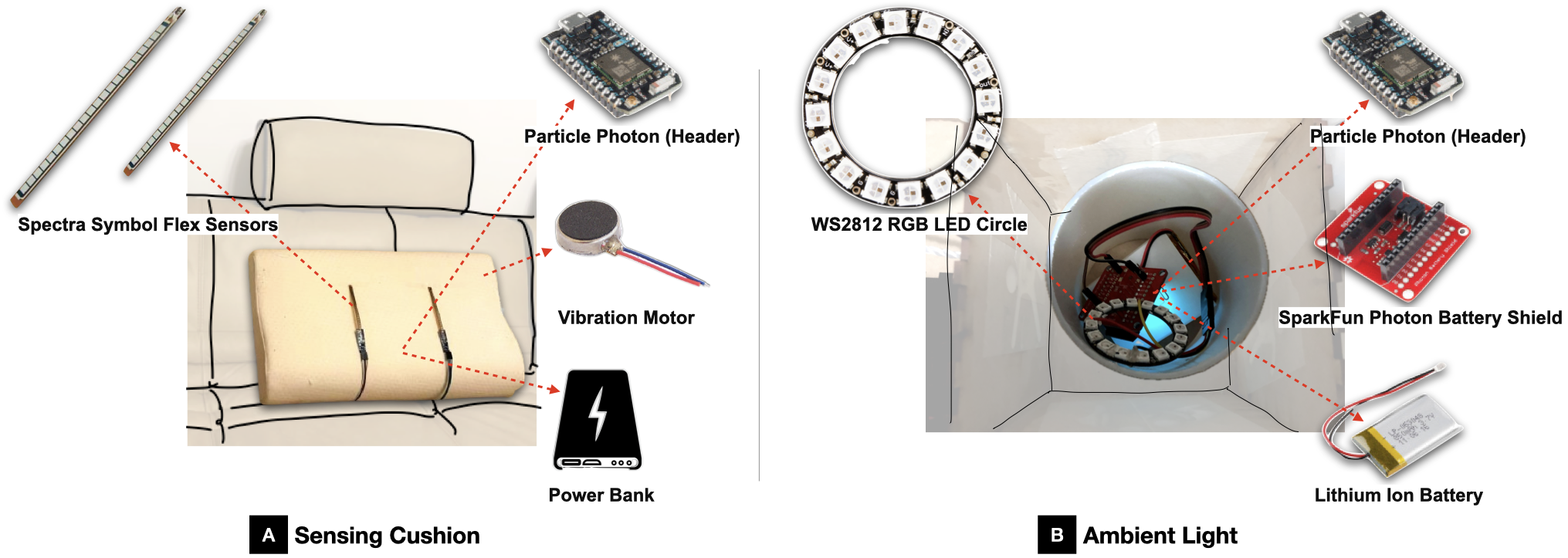}
    \caption{The components of the \textit{CushLumi} system. The \textit{CushLumi} system major used Particle Photon microcontrollers\protect\footnotemark[1], which was programmed in C++, to process real-time data from the flex wiring sensors and control the LED circle. The wireless communication between the hardware components was controlled using Particle's cloud platform. 
    IFTTT\protect\footnotemark[2] was also used to prototype simple time-based and remote control of the system (e.g., muting or unmuting the light via a phone widget or a scheduled applet).}
    \label{fig:components}
\end{figure*}

\begin{figure*}[htbp]
    \centering
    \includegraphics[width=1\linewidth]{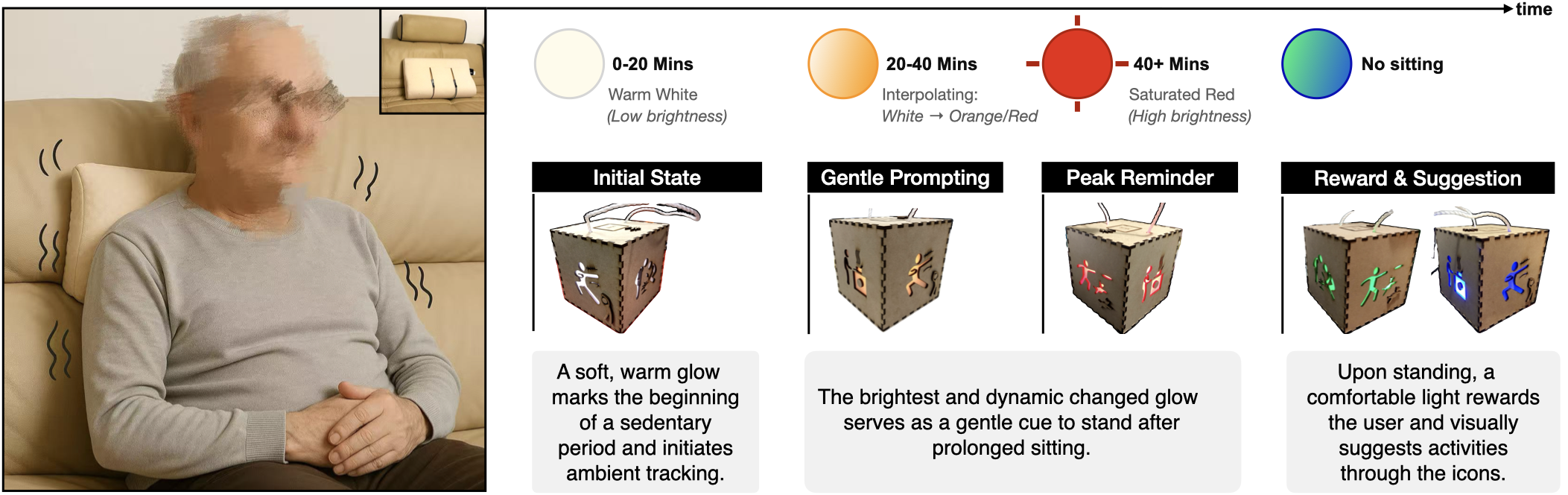}
    \caption{The light behavior logic of the physical data representation in the \textit{CushLumi} system. This figure shows a participant leaning on the Sensing Cushion and examples of the feedback provided by the Ambient Lamp, which changed depending on the duration of time one was sitting or leaning on the cushion. Crafted from warm-toned wood to blend into the home, it translates sedentary time into a gradually brightening ambient glow across three intensity levels. Upon standing, the device does not turn off but instead shines a pleasant color through laser-cut patterns (e.g., a stretching figure), acting as a positive reward and a gentle, suggestive cue for the user’s next activity.}
    \label{fig:highprototype}
\end{figure*}

\autoref{fig:schematicdiagram} presents an overall schematic of the \textit{CushLumi} system, while \autoref{fig:technicalimplementation}, \autoref{fig:components} and \autoref{fig:highprototype} illustrate its layered architecture, hardware components, and light behaviour. In everyday use, the cushion detects when an older adult sits down, stands up, and remains seated for prolonged periods. These events are processed in a simple cloud‑based pipeline that tracks sit durations and drives a small wooden lamp placed nearby. As uninterrupted sitting accumulates, the lamp’s colour and brightness gradually shift from a soft warm white toward a more intense, reddish glow; when the person stands, the lamp briefly plays a colourful celebratory animation before returning to its baseline state. Laser‑cut patterns on the lamp depict positive everyday activities (e.g., stretching, gardening), and a touch‑sensitive top surface serves as a mute button, allowing users to temporarily silence the feedback and thereby maintain a sense of agency. Further design explorations and rationale via RtD, technical details regarding the sensing hardware, light‑mapping logic, and cloud infrastructure are provided in \autoref{appendixRtD} and \autoref{appendixCushLumi}.

\subsection{Procedures}
Herein, we describe the procedures for each phase of our RtD methodology.

\subsubsection{Phase 1}
To understand the context of sedentary behavior and older adults' relationship with technology, we first conducted ethnographic observations of 15 older adults in a local nursing home. We focused on their daily routines, sitting habits, and interactions with everyday objects and technologies. Field notes documented locations, durations, and social contexts of sitting, as well as observed uses of digital devices. Following the observations, we conducted semi-structured interviews with 12 of these older adults. 3 participants were excluded from the interview phase due to scheduling conflicts or health-related fatigue. The interviews covered topics such as daily activity patterns, attitudes toward sedentary behavior, experiences with digital technologies (including smartphones and other screen-based devices), and preferences for how health-related information should be communicated. Interviews also invited participants to reflect on specific situations observed during the ethnographic phase.

During this phase, we introduced early design probes, including the smartphone-based intervention and later the low-fidelity sketch of a cushion that visualizes sitting time through gradually illuminating patterns. 

\footnotetext[1]{Particle Photon microcontrollers: https://www.particle.io/}
\footnotetext[2]{IFTTT: https://ifttt.com/}
\subsubsection{Phase 2}
\begin{figure}[htbp]
    \centering 
    \begin{subfigure}[b]{0.48\textwidth}
        \centering
        \includegraphics[width=\textwidth]{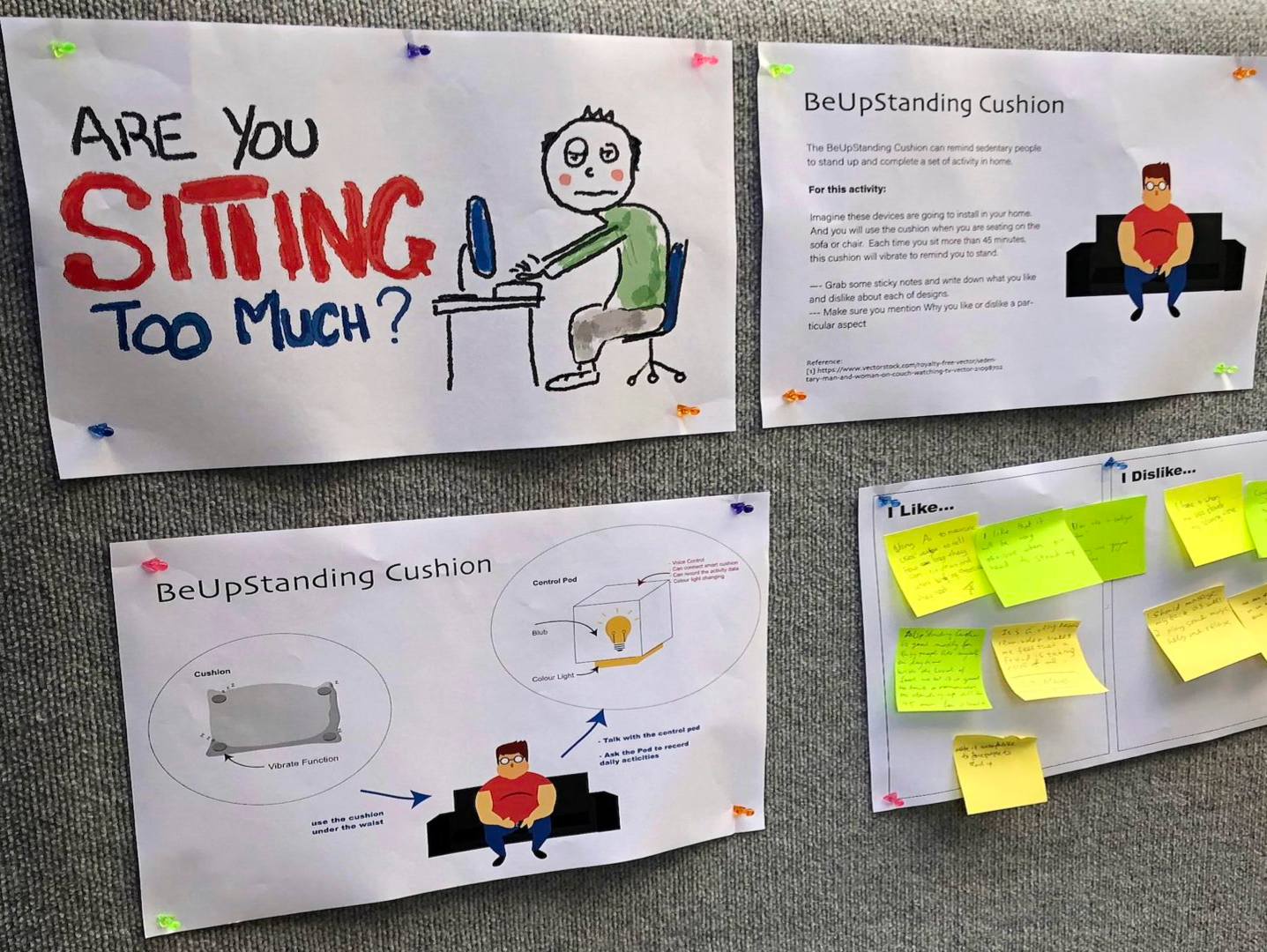}
        \caption{} 
        \label{fig:sub-c} 
    \end{subfigure}
    \hfill 
    \begin{subfigure}[b]{0.48\textwidth}
        \centering
        \includegraphics[width=\textwidth]{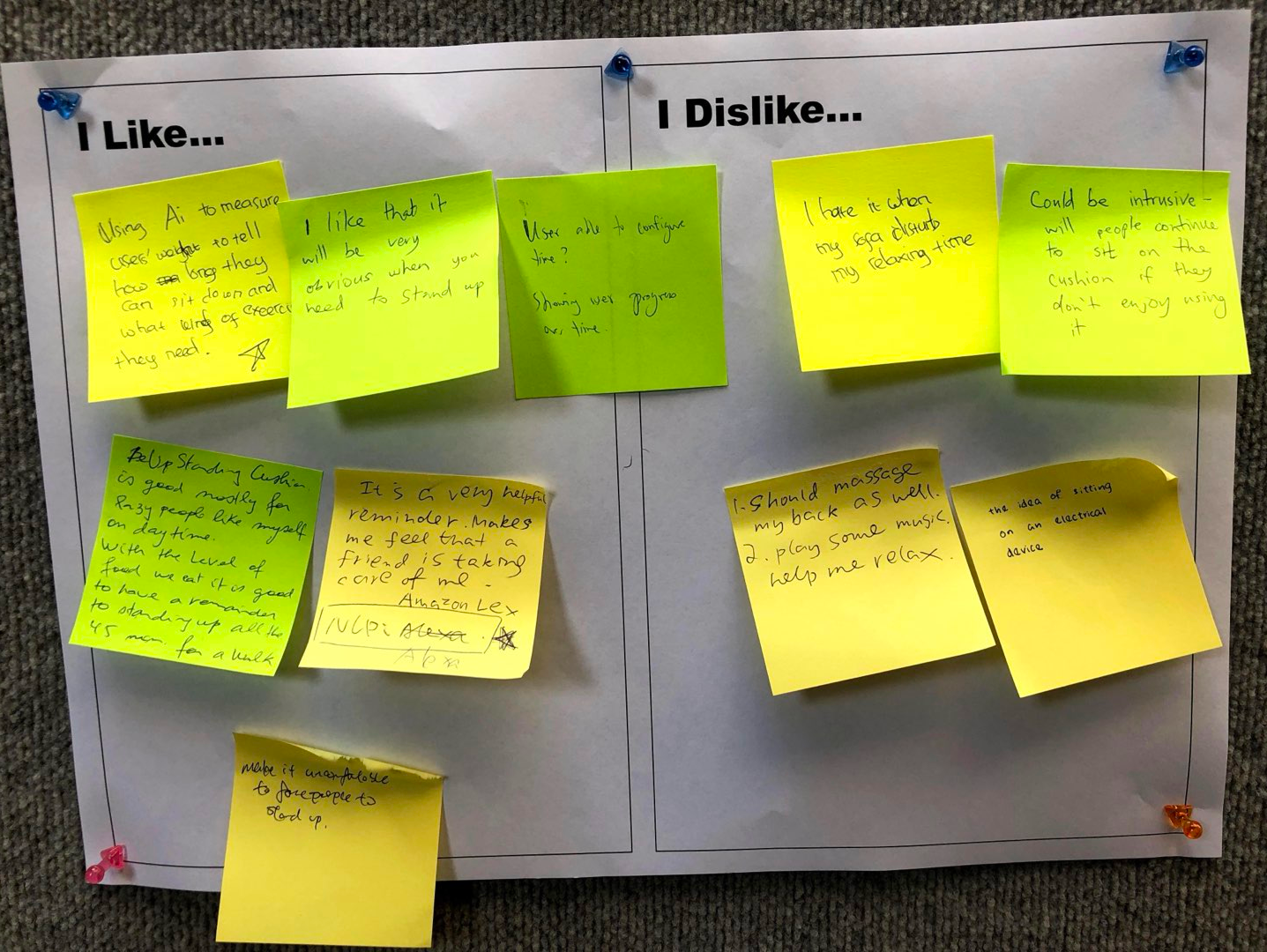}
        \caption{} 
        \label{fig:sub-d} 
    \end{subfigure}
    \caption{The materials used during the Phase 2 poster demo discussion session, where (a) show intervention concepts of potential scenarios and user actions. (a) and (b) also depict user feedback on sticky notes that were later used during affinity diagramming to analyze key themes.}
    \label{fig:demosession}
\end{figure}

Building on the feedback from Phase 1 smartphone complexity and preferences for familiar physical objects like cushions, we refined the concept into a two-part system: a sensor-equipped cushion that would capture sedentary behavior data, and a separate desktop object to display it. Our initial hypothesis for the display relied on relatively explicit visualizations of time, such as a digital bar chart or integration with existing smart home devices. To explore reactions to this concept, we conducted a poster demo discussion session (\autoref{fig:demosession}). We presented visual representations of the proposed system, including sketches and diagrams of the sensor-equipped cushion and the envisioned data display forms. The session was structured as a series of guided discussions around the posters, in which participants were invited to comment on perceived usefulness, clarity, emotional tone, and fit with older adults' everyday lives using sticky notes and conversation.

\subsubsection{Phase 3}
Insights from Phase 2 prompted a shift toward an embodied data narrative that emphasizes gentle awareness rather than clinical monitoring. We used these insights to guide the design and implementation of the \textit{CushLumi} system. To evaluate the system in a real-world context, we deployed the system in the homes of 5 older adults with sedentary habits for one full day. Before deployment, we installed the sensing cushion on a chair the participant frequently used and placed the ambient light box in a visible location in the same room (e.g., on a side table or desk). Participants were instructed to go about their day as usual while the system unobtrusively captured their sitting behavior and provided ambient feedback through the light.

At the end of the day, we conducted semi-structured interviews with each participant in their home to probe their interpretations and feelings towards the data physicalization (e.g., \textit{``When you saw this box light up, what did you think it was telling you?''} and \textit{``How did this feeling compare to a phone alarm?''}). We also asked participants to reflect on how the system fit into their routines, whether and how it influenced their awareness of sitting, and how manageable and comfortable they found the interaction.

We obtained informed consent from all participants before each phase. For older adults in the nursing home and in-home deployments, the study procedures were also coordinated with facility staff and family members when appropriate. Participants could withdraw at any time without penalty, and all collected data was pseudonymized during analysis and for reporting herein.

\subsection{Data Collection and Analysis}
Across all phases, we collected qualitative data through field notes, design artifacts, and audio-recorded interviews. All semi-structured interview audio from Phases 1 and 3 were transcribed verbatim. During Phase 2, we documented the poster demo discussion session through detailed notes and, where permitted, audio recordings that were later summarized. We also retained sketches, posters, and prototype artifacts as part of the RtD documentation. 

We analyzed the semi-structured interview data using reflexive thematic analysis following Braun and Clarke’s approach, as adapted in prior HCI research \cite{lin2019using,bowman2023using}. Two researchers independently open-coded the transcripts, identifying segments related to perceptions of sedentary behavior, reactions to different forms of data representation, emotional responses, and experiences with the prototypes. They then met to discuss and merge their findings into a unified codebook. This codebook was systematically applied to the entire dataset, with the researchers working to refine the codes and group them into higher-level themes. The resulting themes were then related back to the three phases of the RtD process, informing both the iterative design decisions described above and the final set of findings reported herein. Note that we use annotations such as \textit{P3$\_$Phase1} to indicate that, in this example, a quote was from P3 during Phase 1.

\section{Findings}
Our analysis reveals how shifting from explicit numerical tracking to ambient physicalization fundamentally alters older adults' relationship with their health data. We structure these findings to illustrate how atmospheric abstraction can mitigate data anxiety (RQ1) while fostering a sense of user agency through suggestive guidance (RQ2).

\subsection{RQ1: Mitigating Data Anxiety through Atmospheric Abstraction}

\subsubsection{Everyday Interior Design Factors as Low‑burden Sensing Surfaces (Phase 1).}
Through observations and interviews, we found that many participants struggled with smartphones and avoided ``one more digital device''. 
We also found that cushions repeatedly appeared as mundane but central objects in long sitting episodes (living rooms, bedrooms), which have opportunities to highlight the intelligence of a particular data-physicalized thing that can sustain intimate entanglements in everyday settings align with previous work \cite{jansen2015opportunities,frauenberger2019entanglement,Yu2025}. 
Embedding sensing into the cushion promised low additional burden and minimal change to their physical environment.

\subsubsection{Creating a Meaning-Making through Atmosphere (Phase 2).}
Because connecting the choice of material and embodiment to interactions can help tailor a physicalization to a particular target audience \cite{li2025crafting,miao2025physlens}, we refined the concept into a two-part hybrid system: a sensing cushion plus an external display. 

Initially, we tried numeric and bar chart visualizations, assuming they would support clear self-monitoring. But in the poster feedback from both older adults and experts revealed strong negative reactions: graphs felt clinical, judgmental, and reminiscent of hospitals, raising concerns about data anxiety and long‑term engagement. This pushed us towards ambient, non‑numeric feedback using a small, furniture‑like object inspired by \cite{Bae2022,suzuki2017fluxmarker}.

Discussions during this phase had strong reactions to explicit, quantitative health data and to clinical, impersonal visual forms such as bar charts. Participants described some of these representations as \textit{``corny,'' ``boring,'' and ``annoying''} (P15) related such representations to past hospital experiences (e.g., \textit{``It reminded me of how I felt when I needed to go to the hospital''} (P15)) or to a sense of being judged by the data (e.g., feeling \textit{``oppressed''} (P13)). These articulated concerns, together with experts' remarks about potential negative feedback loops for long-term behavior change, informed our subsequent design shift toward more abstract, ambient, and positively framed representations of sedentary behavior.

\subsubsection{Leveraging Abstraction and Feelings of Companionship for Meaning Making (Phase 3)}
A key outcome of the \textit{CushLumi} system was its capacity to convey information without inducing the stress associated with explicit tracking. Participants with negative prior experiences of digital health trackers reported a clear preference for the ambient light. The abstraction of time into a non-numeric glow was central to this effect. P12$\_$Phase1 articulated this contrast: \textit{``That bracelet I had (...) it was like a ticking death-countdown strapped to my wrist, ready to buzz the moment I overstayed my sit—every vibration screaming I’d failed again. This (...) doesn’t count me down to doom. If it flares bright, it’s just a fact, like the kettle boiling. No verdict. I can glance, think `ah, okay,' and let it go.''}. Reframing the light as a neutral environmental signal, rather than a personal performance metric, was fundamental to mitigating P12's anxiety. 

By moving beyond mere abstractions, we were able to craft an entity that participants experienced not as a source of information, but as a relational companion dwelling within their domestic environment. Participants consistently interpreted the system’s light as a social actor exhibiting a gesture of care.  P3$\_$Phase1 drew a direct contrast with being reprimanded: \textit{``It’s not telling me off. It’s like a friend saying, ‘Shall we stretch a bit?’''}. This reframing was particularly effective for individuals who had previously rejected numerical trackers. The participants who found charts and numbers stressful now described the light with positive, social language, calling it ``gentle,'' ``non-clinical,'' and ``like a pet on the table''. This shift from a clinical to a social frame removed the feeling of being under surveillance. As P1$\_$Phase1 explained: \textit{``I didn’t feel judged. It’s just a glow. It doesn’t make me feel like I’m in a check-up.''}. This positive emotional reception was corroborated by our observation notes, which recorded 11 instances of participants smiling upon noticing the light, and zero instances of frowning or verbal resistance. In two cases, participants greeted the light verbally (e.g., \textit{``Okay, I see you''}), suggesting a para-social relationship rather than a purely utilitarian interaction.

This sense of companionship was achieved by changing the room's feeling. Participants perceived the device not as an information display but as felt it could integrate part of their home's ambiance, providing a canvas upon which they could construct personal significance. P9$\_$Phase3 said \textit{``It is not data, it is like an atmosphere, like when the sun sets, and the room gets warm and golden, you know the day is ending. This light is like that. It doesn't count on me. It just changes the feeling of the room. This change in lighting is very romantic ... a very gentle way. Its a very poetic way of knowing time has passed.''}.  By comparing the light to a sunset, the participant was not merely having an experience, he was actively interpreting the abstract data by connecting it to a familiar, natural phenomenon. This act of drawing an analogy transformed the device from an information display into a part of the room's narrative. This demonstrates that for this user group, data accessibility is contingent not only on simplification but also on its integration into a form that allows for personal meaning-making.

\subsection{RQ2: Enabling User Agency and Suggestive Guidance}
Our second research question explored how this ambient approach differs from traditional digital reminders. We found that the \textit{CushLumi} fostered a different relationship with the user by acting as a nudge rather than a nag. This was achieved by functioning as a non-interruptive suggestion, providing users with ultimate control, and guiding action by presenting possibilities rather than prescribing solutions. 

\subsubsection{Subtly Guiding User Behavior Rather Than Issuing Orders (Phase 3)}
During the Phase 3 deployment, 12 standing events were observed across five participants. In 8 of these 12 standing events, the participant’s first activity after standing corresponded to one of the icons on the side of the lamp (i.e., going to the kitchen to boil water, taking the dog outside to play, going outside to garden, or stretching the body). P9$\_$Phase3 noted that upon seeing the stretching side of the lamp light up, he found himself mirroring the action without conscious thought. P4 was also observed mirroring the posture shown in the pattern, explaining \textit{``I saw the little man stretching, so I did the same. It felt natural.''}. Unlike traditional alerts that function as a command to be obeyed, the system employed a non-linguistic, mimetic form of behavioral guidance that felt self-directed, rather than like following an order.

While traditional reminders typically end their interaction once the target behavior is achieved, the \textit{CushLumi} used this moment to open a space for action. P11$\_$Phase3, a gardener, described that \textit{``When I stand up and it shines that green light through the little flower pattern ... it's like a reward. It's not just telling me good job for standing, it's reminding me of something I love. It turns a chore, getting up, into a happy thought''. }. For this participant, the system connected a health-oriented action (standing) to a source of intrinsic motivation (gardening), reframing the act from a chore into a positive opportunity. As the system’s ambient cues were perceived as polite and suggestive, the light created a space for self-directed thought. As P4$\_$Phase3 described \textit{``I didn’t feel pushed. I just thought, ‘Maybe I’ll water the plants’ and I did.''}. 
This shift from a need for compliance to inspiration was facilitated by the gradual brightness being interpreted as a gentle buildup and the laser-cut icons serving as implicit, embodied suggestions for post-standing activities.

\subsubsection{Promoting Acceptance through Form, Flexibility, and User Control (Phase 3)}
Unlike typical alerts that demand immediate attention, the \textit{CushLumi} was perceived as non-interruptive. This allowed users to respond at their own pace, integrating the prompt into the natural rhythm of their day. P8$\_$Phase1, who spent her afternoons watching television, reflected: \textit{``My daughter's phone calls are an interruption. The doorbell is an interruption. This light ... it's a suggestion. It waits for a natural break, like the TV ads. It never yells at me. It's polite''.} 

The primary factor driving user acceptance was the system’s support for user agency. The touch-sensitive mute button served as a foundation for user trust and acceptance, particularly for skeptical participants. P14$\_$Phase3 described how \textit{``The first thing I did was touch the top to see if I could turn it off. I could, that was the most important part because I knew I was in charge ...if I couldn't turn it off, it would have been unplugged in five minutes.''}. This sense of agency transformed the device from a potential intrusion into a tool the participant was willing to engage with. 

We found that the mute function was used by three participants as a way to have a dialogue with the device. P3$\_$Phase2 articulated \textit{``I liked that I could silence it, it’s like saying, I hear you but not now.''}. This contrasts with persistent digital alerts that do not listen, as P5 noted in Phase 1. 
We also found the acceptance was cemented by the system's non-stigmatizing form factor. Participants repeatedly emphasized that it did not look like a medical device, which allowed it to blend seamlessly into their homes and lives (e.g., \textit{``It’s just a lamp. I like that it doesn’t scream `health' at me.''} P2$\_$Phase1). This suggests that the combination of a familiar form, interpretive flexibility, and tangible user control was central to the system's potential long-term acceptability, differentiating it from paternalistic technologies and affirming the autonomy of its users.

\section{Discussion}
Our study aimed to explore how to make sedentary data accessible for older adults and how this method differs from traditional reminders. The results show that the key to accessibility is not to simplify quantitative data but to transform it into an atmospheric abstraction \cite{10.1145/3706598.3714302, Heijboer2016Peripheral}. Participants reported experiencing the light as gentle, non-clinical, friendly, and as an ordinary lamp that did not make them feel judged or subjected to a medical check-up. This is in line with prior work that argues for supportive instead of paternalistic health technologies for older adults \cite{Yu2025,voinea2024paternalistic}. In this sense, our findings showing that supportive and atmospheric abstractions can be maintained in everyday in-situ use, rather than only in short-term or laboratory settings. Herein, we reflect on our findings and describe their implications for future interventions that support older adults living, and interacting with, their personal data.

\subsection{Rethinking Forms of Physical Data Representations for Older Adults}
In light of our results, it is necessary to shift the model of interaction from one of command and compliance to one of suggestion and negotiation, which better protects a user’s sense of autonomy. Because the system did not provide precise numbers, it avoided generating feelings of being judged and monitored. Participants thus tended to perceive the device as a gentle and companion-like presence \cite{Liu2023VoiceAssistantsCompanionship}. Such positive emotional connections are difficult for traditional health tracking devices to establish, even though they can make health interventions feel more humane and could increase the likelihood they are accepted by users in the long term. 

The ambiguity and gentleness of such designs may have potential negative impacts as well. If a design is too focused on being non-intrusive, it may struggle to achieve its original health intervention goals \cite{Frissen2024ProgressFeedback, Ntsweng2025AALLessons, Lee2023AmbientToDisruptive}. When users ignore a reminder, it is also difficult to determine if this was due to an empowered, autonomous choice or if a design had failed. Rather than trying to find a perfect balance between clarity and ambiguity, it might be more prudent to actively manage this uncertainty and explicitly give final control to the user \cite{10.1145/3532106.3533471}. Our mute button allowed users to negotiate with the system and acknowledge a reminder while choosing to act on it later. Allowing users to negotiate, delay, or selectively decline prompts may help reduce feelings of being controlled or judged and may lower reactance and data-related anxiety \mbox{\cite{thudt2018self,ramesh2024exploratory}}. It can also help preserve a sense of autonomy and dignity, which prior work has identified as a central concern in the design of technologies for aging populations \mbox{\cite{li2025crafting,lupton2017feeling}}. In line with the exploratory, in-situ nature of our RtD process, we suggest that such negotiated engagement should be seen as an informative outcome that complements more conventional measures of short-term effectiveness.

\subsection{Leveraging Research Through Design for Exploratory Digital-to-Physical Translations}
Our primary research challenge was how to translate an abstract digital concept, sedentary time, into a meaningful physical experience. Our initial design concepts gravitated towards literal translations, envisioning physical versions of bar charts or glowing orbs that simply mimicked on-screen dashboards. We quickly realized these early concepts were either too complex, too distracting, or would fail to evoke gentle awareness. These realizations highlighted the unique strength of RtD approaches. Rather than being locked into a flawed initial plan, we used the artifacts themselves as our primary mode of inquiry, allowing us to rapidly adapt and learn from successive design concepts and prototypes.

This experience thus suggests that for this class of problem, RtD can act as a form of experiential inquiry. As our central question was phenomenal rather than technical (i.e., what does it feel like to live with an ambient representation of your own data?), we needed to build artifacts and have end users live with them \cite{10.1145/291224.291235, 10.5555/1841406}. Through this process, we uncovered the direct conflict between a desire for calm ambiance and the need for a clear, actionable alert \cite{Lee2023AmbientToDisruptive, Pousman2006AmbientTaxonomy}. Our artifacts became physical arguments, each one testing a different stance on this conflict.

\subsection{Living with Data through Ambient Co-dwelling}
% \subsection{Living with Data between Ambient Abstraction and Joint Action}

Our investigation suggests that making sedentary data accessible for older adults requires more than merely simplifying numerical visualizations. It demands a fundamental shift in the user-information relationship—a concept we frame as living with data. To live with data implies that personal health information is no longer treated as an external metric requiring active, focused interrogation. Instead, it becomes an integral, unobtrusive quality of the domestic environment \cite{Kaziunas02012018}. In this paradigm, the goal shifts from the rigorous tracking of performance to the cultivation of a sustained awareness\cite{vandeweerd2020homesense, vanhoof2011ageing}, where data is woven into the fabric of daily life rather than standing apart as a distinct object of scrutiny.

This state of living with data is realized through the mechanism of ambient co-dwelling, where quantitative data is transformed into an atmospheric abstraction \cite{10.1145/3706598.3714302,Ntsweng2025AALLessons}. Unlike traditional trackers that participants described as ticking death-countdowns (P12), the ambient light allowed data to dwell within the home as a gentle and friendly presence, akin to a sunset or a household pet. By integrating information into the feeling of the room (P9), the system shifts the experience from surveillance to co-dwelling. This approach enables users to engage with their health status through peripheral perception \cite{weiser1996calm, Heijboer2016Peripheral, Bakker2016PeripheralInteraction}, allowing the data to exist comfortably in the background without demanding immediate attention or inducing judgment.

However, this prioritization of ambient comfort reveals a complex tension between living with data and the mechanics of joint action motivation \cite{10.1145/3643834.3660689}. While the physicalization of sedentary behavior could mitigate the anxiety associated with surveillance, the resulting abstraction could inadvertently obscure the precise behavioral cues required to trigger synchronous social interventions \cite{dourish2001where, schmidt2002awareness}. For older adults in everyday life, the very features that make the system unobtrusive simultaneously dampen the immediate social pressure to act together \cite{rogers2006moving, kotary1995age}. Consequently, physicalization approaches might navigate a delicate trade-off where the aesthetic integration of health information fosters long-term acceptance but may reduce the explicit urgency typically needed to drive mutual adherence and immediate behavioral change.

\subsection{Design Implications}
Our findings can be synthesized into three design implications to support older adults living with their own data. First, sedentary histories should be continuously folded back into a physical model rather than appearing as isolated readouts. This was evidenced by our design decision to separate health data sensing from display, which moved data away from screens and into the background of the home. Second, personal sedentary data should leverage subtle modulations of light and metaphors rather than explicit numbers, thresholds, or performance goals so older adults can perceive changes in their sedentary patterns. Third, because older adults often associate feelings of being monitored or judged with dashboard-style visualizations and surveillance-oriented gerontechnology, quiet and negotiable forms of engagement should be used to preserve the richness of one's personal health data. 

These implications outline a design direction for future tangible health technologies that shifts the emphasis from monitoring and compliance to empowerment, allowing older adults to maintain dignity and autonomy while data are woven into their everyday lives. In the context of health support, enabling older adults to reduce reactance, avoid data anxiety, and make it more likely that the intervention will be lived with over time.

\subsection{Limitations and Future Work}
While providing novel insights, our research does have several limitations. First, the one-day deployment of the \textit{CushLumi} system during Phase 3 prevented us from understanding the potential long-term effects of the intervention, such as what occurs when its novelty wears off or what would occur if it were utilized outside a controlled, in-home setting. Our future plans include a longitudinal, in-situ study lasting several months to examine whether and how the atmospheric abstraction supports sustained changes in sedentary patterns and ongoing engagement over time. For instance, we are interested in understanding the degree to which atmospheric abstraction can lead to sustained behaviour change, and how users’ interpretations of the ambient light shift with prolonged use. 
It is important to note that evaluating the clinical efficacy or long-term health outcomes of the device was outside the scope of this exploratory design study, which focused on user experience and acceptance.

The small sample size of five older adults in Phase 3, also limits the generalizability of our findings to the broader older adult population and to other cultural or geographic contexts. As our research was conducted with older adults in an Australian Western context, future work must examine how ambient data physicalizations are perceived in other cultural settings, including non-Western contexts that may have different norms around aging, health, family dynamics, and family care.

Objective studies of sedentary behavior also typically rely on accelerometers within wearable devices to measure sedentary time \cite{healy2008objectively,healy2008breaks}. Our cushion-based sensing plays a similar role, but within a piece of decor. We did consider the use of load cells, pressure mats, and camera-based sensing to measure sedentary time, however, load cells and mats required more rigid structures and careful calibration, while cameras raised privacy concerns. Flexible sensors offered a good trade-off between simplicity, comfort, and ease of replication. It would, however, be worthwhile to explore other sensing methods and form factors in the future.

Architecturally, the \textit{CushLumi} system was also designed so that participants’ own sensor data could serve as a basis for secondary modelling and personalisation. This adaptive layer was not activated in the present study but will be explored in future work to better understand the well-being data. Furthermore, we plan to iterate on the \textit{CushLumi} system's design by exploring different materials, forms, and sensory modalities (e.g., subtle temperature changes) to expand the vocabulary of atmospheric physicalizations that are possible.

\section{Conclusion}
Using a three-phase Research through Design methodology, our research explored how sedentary data could be translated into an atmospheric, ambient physicalization for older adults. By offering gentle, suggestive guidance rather than intrusive commands, participants' data anxiety was mitigated and older adults felt empowered to be less sedentary. Our \textit{CushLumi} System prototype fostered positive, relational engagement with one's personal health data, marking a shift from merely informing users to enabling them to live reflectively with their data.

%%
%% The acknowledgments section is defined using the "acks" environment
%% (and NOT an unnumbered section). This ensures the proper
%% identification of the section in the article metadata, and the
%% consistent spelling of the heading.
\begin{acks}
This research was undertaken while I was a student at The University of Queensland, and I gratefully acknowledge the support and facilities provided by the university. I also extend my sincere thanks to the older adults who participated in this study for their valuable contributions.
\end{acks}

%%
%% The next two lines define the bibliography style to be used, and
%% the bibliography file.
\bibliographystyle{ACM-Reference-Format}
\bibliography{chi26-1682}

%%
%% If your work has an appendix, this is the place to put it.
% \clearpage
\clearpage 
\appendix
\section{Design Explorations and Rationale via RtD}
\label{appendixRtD}
We initially explored the use of a smartphone application that aimed to deliver sedentary behavior feedback using conventional interaction methods and visualizations. During Phase 1, however, participants had an aversion to complex interfaces, small text, and the cognitive load of managing yet another app. For example, one participant remarked, \textit{``If I can choose, I don't want to use smart phone, it has too many functions''} (P8). Participants also explicitly stated that they wanted to avoid reminders that felt like nagging. This feedback, in addition to prior work that identified that longer sedentary periods are associated with higher clustered metabolic risk and all‑cause and cardiovascular mortality \cite{owen2010sedentary,healy2008objectively,healy2008breaks} and recommendations that emphasized interrupting sitting every 30 to 60 minutes \cite{owen2010sedentary,healy2008breaks,martin2015interventions,https://doi.org/10.1111/obr.12215}, led to the exploration of interventions that could be embedded within existing physical environments and daily routines to discourage sedentary behavior.

As participants in Phase 1 identified cushions as one of the most common artifacts they sat on (e.g., \textit{ ``This [cushion] is something I always use when I sit here.''} P1), we created a design concept sketch of a cushion that would visualize one's sitting time through gradually illuminating patterns on its surface. This early concept served as a probe during Phase 2 to investigate the feasibility and acceptability of embedding data feedback in a familiar physical object.

Building on the feedback obtained during Phase 2, we then designed and implemented the \emph{\textit{CushLumi} System} (\autoref{fig:schematicdiagram}). 
Our design included a cushion and a light because prior ambient display work leveraged light for slowly changing, low priority information\cite{wisneski1998ambient,7927391,mclaughlin2020designing,feine2025integrating} and research on aging in place environments highlighted furniture (e.g., lighting) as a key interior design factor for supporting sleep, mood, and daily functioning in older adults \cite{Yu2025}.

To make the connection between the formative findings and the final form factors explicit, we summarize our design explorations across the three RtD phases in \autoref{tab:design_rationale}. For each phase, we articulate the design goals, the rationale grounded in user and expert feedback, the configuration we chose to pursue next, and representative participant quotes that directly motivated these choices. 

\section{\textit{CushLumi} System Implementation}
\label{appendixCushLumi}
The system consisted three parts: a \textit{Sensing Cushion} and an \textit{Ambient Light} lamp, which are the two physical artefacts users directly engage with, and a cloud‑based processing component that acts as an invisible bridge, receiving sit/stand events from the cushion and driving the lamp’s gradual light changes and celebratory animations. Whenever a user sat on the cushion, its deformation was detected and the system began tracking the duration of one's sitting. This information was then used to change the lamp's hue and light intensity to motivate the user to stand up and move around.

The \textit{Sensing Cushion} contained two Spectra Symbol 4.5-inch flexible sensors that were embedded in the foam of a 30 cm × 60 cm foam cushion. The sensors formed a shallow V pattern that was intended to roughly align with a user's thighs and pelvis. Each flex sensor was sampled by a Particle Photon microcontroller at 10 Hz. The microcontroller read both analog channels, detected bending due to body weight, and aggregated readings over a 3 second window with hysteresis.

The \textit{Ambient Light} was a small wooden box that featured laser-cut patterns of positive activities (e.g., stretching, gardening) on its faces, giving it the appearance of a tabletop lamp. The lamp measured approximately 8cm × 8cm ×8 cm and was powered by a single-cell LiPo battery, allowing it to be freely placed on side tables or shelves. The box contained 16 addressable WS2812 RGB LED Circle that glowed depending on one's sitting time (i.e., 0 - 20 minutes -> no intervention needed so the LEDs were warm white, at 20 - 30 percent brightness; 20 - 40 minutes -> gentle prompting where the hue linearly interpolated from warm white to orange/red as a function of the elapsed minutes, with the brightness increasing moderately over time; 40+ minutes -> standing is recommended so the hue shifted toward saturated red, with the brightness increased to 70 to 80 percent, with smaller additional increases if the user did not stand up). In line with work showing that even replacing small amounts of sedentary time with light intensity movement can benefit metabolic health \cite{healy2008objectively,healy2008breaks}, the mapping was intended to gently nudge users toward more frequent, small breaks instead of signaling a sudden point where it was already too late to see benefits of standing. The intensity levels conservatively operationalized the 30 to 60 minute recommendations reported in sedentary behavior research \cite{owen2010sedentary,healy2008breaks,dempsey2020new}.
When a user stood up, a short celebratory animation was played (i.e., the lamp cycled through rainbow hues for 2 to 3 seconds). The lamp's LEDs were updated at a frequency of 1Hz. A touch-sensitive area on the top served as a Mute button, providing direct control over the feedback and reinforcing the user's sense of agency. Five \textit{CushLumi} Systems were built for use in Phase 3.

\begin{table*}[t]
\caption{Design goals, rationale, and example quotes across three RtD phases}
\label{tab:design_rationale}
\small
\centering
\begin{tabular}{p{0.19\textwidth}|p{0.24\textwidth}|p{0.24\textwidth}|p{0.24\textwidth}}
\toprule[1.2pt]
\textbf{Phase \& configuration} & \textbf{Design goals} & \textbf{Design rationale} & \textbf{Representative quotes} \\
\midrule[1.2pt]

% ---------- Phase 1 ----------
\textbf{Phase 1:} \newline
Formative research, \newline
cushion as a form factor
&
Understand older adults' sedentary routines and attitudes to technology; explore initial ways to surface sitting time without disrupting everyday practices.
&
Through observations and interviews, we found that many participants struggled with smartphones and avoided ``one more digital device''. 

We also found that cushions repeatedly appeared as mundane but central objects in long sitting episodes (living rooms, bedrooms), which have opportunities to highlight the intelligence of a particular data-physicalized thing that can sustain intimate entanglements in everyday settings align with previous work \cite{jansen2015opportunities,frauenberger2019entanglement,Yu2025}. 
Embedding sensing into the cushion promised low additional burden and minimal change to their physical environment.
&
\begin{itemize}[leftmargin=*]\setlength\itemsep{0pt}
  \item \textit{``If I can choose, I don't want to use smart phone, it has too many functions.''} (P8)
  \item \textit{[On cushions] ``This is something I always use when I sit here.''} (P1, field notes)
\end{itemize}
\\ \cmidrule{1-4}

% ---------- Phase 2 ----------
\textbf{Phase 2:} \newline
Poster demo, \newline
separate display object
&
Clarify how sedentary data should be represented; test reactions to explicit quantitative displays; maintain comfort by separating sensing from display.
&
Because connecting the choice of material and embodiment to interactions can help tailor a physicalization to a particular target audience \cite{li2025crafting,miao2025physlens}, we refined the concept into a two-part hybrid system: a sensing cushion plus an external display. 
Initially, we tried numeric and bar-chart visualizations, assuming they would support clear self-monitoring. But in the poster feedback from both older adults and experts revealed strong negative reactions: graphs felt clinical, judgmental, and reminiscent of hospitals, raising concerns about data anxiety and long-term engagement. This pushed us towards ambient, non-numeric feedback using a small, furniture-like object inspired by \cite{Bae2022,suzuki2017fluxmarker}.
&
\begin{itemize}[leftmargin=*]\setlength\itemsep{0pt}
  \item \textit{``It reminded me of how I felt when I needed to go to the hospital.''} (P15, on bar charts)
  \item \textit{``Corny'', ``boring'', and ``annoying''.} (P15, on group discussion)
  \item \textit{``I feel a bit oppressed by it.''} (P13, on explicit numeric feedback)
  \item \textit{It's not just telling me good job for standing, it's reminding me of something I love.}(P11, on explicit numeric feedback)
\end{itemize}
\\ \cmidrule{1-4}

% ---------- Phase 3 ----------
\textbf{Phase 3:} \newline
\textit{CushLumi} System, \newline
field deployment
&
Realise a fully functional system that embodies an ``ambient, gentle'' data narrative; keep sensing invisible while offering legible but non-intrusive feedback; support user agency and positive associations with movement.
&
In order to build the composite relation with a data-physicalized thing in terms of non-neutrality, mutual shaping, and shared intentionalities, we implemented the \textit{CushLumi} system: the cushion stays visually unchanged and embeds flex sensors plus a microcontroller to detect sitting \cite{zhong2025investigating}.

We aim to build and deploy the technological data-physicalized thing as it has intentionality in assembling the cohabiting human beings and other things in the situated context of everyday \cite{wakkary2021things}. At the same time, we also hope that through the cube's form design, we can offer target users more activity suggestions after standing, and motivate users to stand up proactively through intuitive visual cues \cite{anderson2011seductive}.
A small wooden ``ambient light'' box sits on nearby furniture and gradually brightens as sitting accumulates; when the person stands, it shifts to a pleasant colour shining through laser-cut patterns of everyday activities (e.g., stretching, gardening), gently suggesting alternatives to prolonged sitting. 
A touch-sensitive area lets users mute feedback, reinforcing a sense of control.
&
\begin{itemize}[leftmargin=*]\setlength\itemsep{0pt}
  \item Participants often described the box as ``a little lamp'' or ``a decoration'', not as a life countdown reminder.
  \item \textit{``I didn’t feel pushed. I just thought, ‘Maybe I’ll water the plants.’ And I did.''} (P4, deployment interview)
\end{itemize}
\\
\bottomrule[1.2pt]
\end{tabular}
\end{table*}

\end{document}